# A Novel Application of Lifting Scheme for Multiresolution Correlation of Complex Radar Signals

Chinmoy Bhattacharya, *Member, IEEE* and P.R.Mahapatra

*Abstract*— The lifting scheme of discrete wavelet transform (DWT) is now quite well established as an efficient technique for image compression, and has been incorporated into the JPEG2000 standards. However, the potential of the lifting scheme has not been exploited in the context of correlation-based processing, such as encountered in radar applications. This paper presents a complete and consistent framework for the application of DWT for correlation of complex signals. In particular, lifting scheme factorization of biorthogonal filterbanks is carried out in dual analysis basis spaces for multiresolution correlation of complex radar signals in the DWT domain only. A causal formulation of lifting for orthogonal filterbank is also developed. The resulting parallel algorithms and consequent saving of computational effort are briefly dealt with.

*Index Terms*— Biorthogonal, correlation, discrete wavelet transform (DWT), lifting, multiresolution analysis (MRA), radar signal processing.

## I. INTRODUCTION

DISCRETE wavelet transform (DWT) has emerged as a powerful technique in diverse areas such as multiresolution analysis (MRA), compression, detection, matching, and feature extraction of signals and images. DWT, originally implemented through Mallat's filterbank algorithm [8], has been rendered more efficient by the development of the *lifting scheme* that has been incorporated in the new image compression standard, JPEG 2000 [3]. The major advantages of the lifting scheme are (1) flexible factorization of the 2 - channel orthogonal, biorthogonal filterbanks [11], [29] and (2) robustness against using finite precision and integer arithmetic in factorizing the forward transform, with exactly reversible reconstruction [9], [10], [25]. These enable *parallel*, *in place* and *faster* computation of DWT [11], [12]. Several approaches have evolved for hardware implementation of the lifting scheme using dedicated ICs or field-programmable gate arrays (FPGA) [4]-[7].

A number of important signal processing applications, such as matched filtering in radar systems, involve correlation of two complex signals. Multiresolution correlation has distinct advantages over single-resolution radar processing by FFT, e.g. achievement of significant performance gain in the case of synthetic aperture radar (SAR) [13], [14]. An interesting approach to multiresolution detection is devising matching wavelets to a specified signal, i.e., interpreting DWT as a bank of matched filters [20]. A desired signal is detected by a sharper peak response to a specified wavelet filter matched to the signal in question.

Although the lifting scheme has been well developed in the image compression context, its power has not been adequately exploited for correlation of radar signals. This is because of the inherent differences between compression and correlation processes. Evolving a complete and consistent framework for application of DWT by lifting to correlation of complex signals is the subject of this paper.

The primary difference between signal compression and correlation in the DWT domain is that the former involves both analysis and synthesis while the latter requires analysis only [21]. Factorization of the biorthogonal filterbanks into lifting steps *only* in dual analysis bases, as required for multiresolution correlation, is a major development in the current work.

Another major difficulty in performing multiple scale signal correlation by DWT is the shift-variant nature of *critically sampled* DWT. The downsampling factor of two in the 2 - channel filterbank of forward DWT introduces periodic shift-variance at the output due to dyadic change of time scale [1], [2]. Shift-variance is known to cause problems in multiresolution image representation and texture analysis [15]. MRA is therefore often implemented in the undecimated DWT domain. In [16], [17] speckle reduction in SAR images is achieved by redundant use of 2 - channel wavelet filterbanks. Other approaches include progressive correlation in hierarchical scales using both DWT and FFT [18] and near shift-invariant approaches such as complex wavelet transform (CWT) [19], [31].

It has been demonstrated recently that multiresolution correlation between complex signal vectors can be achieved by using the shift circulant property of DWT coefficients for one dyadic scale change [21]. However, this involves redundant DWT of the matching signal to determine even and odd shifts in the received signal, and vector multiplication with the stored shift circulant DWT matrix, at each scale. The advantage gained is *exact* shift- and scale-invariant correlation between complex signals simultaneously at dyadic scales. This approach forms the background for the lifting scheme formulation for correlation developed in the current paper.

In the next section we provide a brief overview of correlation by shift-invariant DWT (SIDWT) in biorthogonal basis spaces. In Section III, after a short exposition on the paraunitary property of biorthogonal polyphase matrices we derive a framework for correlation in the polyphase domain. The lifting scheme factorization for orthogonal, causal form of Daubechies4 filter and biorthogonal (5/3), (9/7) filters in dual analysis bases are developed for SIDWT based correlation. Multiresolution correlation of complex radar signals at dyadic scales using the lifting schemes is dealt with in Section IV. Section V briefly touches upon the computational issues resulting from the inherent parallelism of the algorithm. Finally, in section VI we draw the major conclusions of the paper.

*Notations*: $x[n]$ and $x(z)$ are vectors in time and $z$-domains respectively, i.e. $\tilde{h}[-n] \Leftrightarrow \tilde{h}(z^{-1})$. The $\sim$ notation is for analysis. $P_j, Q_j$ are projection operators and $W_T$ is the transform domain matrix operator; the symbols $\circ$ and $\otimes$ stand for circular correlation and convolution operators respectively. Matrices are represented as, e.g. $P(z)$, $Z$ is the set of integer numbers. FIR filters are given by Laurent polynomials, i.e.

$$h(z) = \sum_{n=k_1}^{k_2} h[n] z^{-n}, \{k_1, k_2\} \in Z ,$$

where the degree of the polynomial is $|k_2 - k_1|$.

## II. REVIEW OF THE CORRELATION ALGORITHM BY SIDWT

Consider *finite*, wide order stationary sequences as is common in radar applications. For a real vector $x_j[n]$ of length $L = 2^J$, sampled uniformly over $n \in Z_+$, the $\ell_2$ norm of $x_j[n]$ is

$$\|x_j[n]\| = \sqrt{\sum_{n=0}^{L-1} x_j^2[n]} < \infty . \quad (1)$$

The approximation and detail components of $x_{(j-1)}[n] \in V_{(j-1)}$ in dyadic, coarser multiresolution analysis spaces $\{V_j, W_j\}$ spanned by the basis vectors, respectively $f_{j,k}[n]$ and $y_{j,k}[n]$, are

$$P_j x_{j-1}[n] = \sum_k <x_{j-1}, \tilde{f}_{j,k}> f_{j,k}[n] = \sum_k \tilde{a}_{jk} f_{j,k}[n] \in V_j$$
$$Q_j x_{j-1}[n] = \sum_k <x_{j-1}, \tilde{y}_{j,k}> y_{j,k}[n] = \sum_k \tilde{b}_{jk} y_{j,k}[n] \in W_j. \quad (2)$$

Here $\{\tilde{a}_{jk}, \tilde{b}_{jk}\}$ are the approximation and detail DWT projection coefficients derived in the dual biorthogonal analysis spaces $\tilde{V}_j$ and $\tilde{W}_j$ spanned by dual basis vectors $\tilde{f}_{j,k}[n]$ and $\tilde{y}_{j,k}[n]$ respectively. For MRA at dyadic scales $2^1 \le 2^j \le 2^J$, the synthesis of vectors $\{x[n], y[n]\} \in V_0$ in dual biorthogonal basis spaces are given by,

$$x[n] = \sum_j \sum_k \tilde{b}_{jk} y_{j,k}[n] + \sum_k \tilde{a}_{Jk} f_{J,k}[n]$$
$$y[n] = \sum_j \sum_k d_{jk} \tilde{y}_{j,k}[n] + \sum_k c_{Jk} \tilde{f}_{J,k}[n] \quad (3)$$

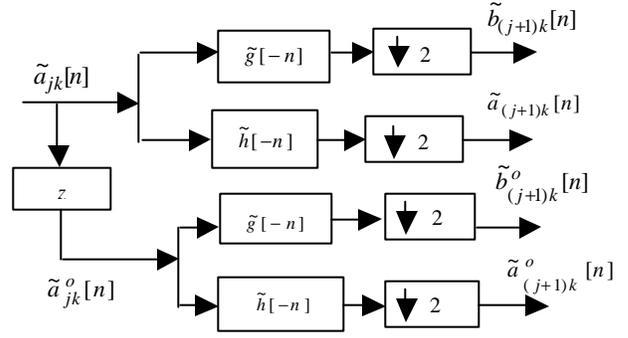

Fig. 1. Filterbank schematic for SIDWT decomposition.

where, from (2),

$$P_J x_{J-1}[n] = \sum_k \tilde{a}_{Jk} f_{J,k}[n], P_J y_{J-1}[n] = \sum_k c_{Jk} \tilde{f}_{J,k}[n]$$

are the projections of $\{x[n], y[n]\}$ at the coarsest scale $2^J$. Due to the biorthogonal property of dual basis spaces [2], it follows from (2) and (3) that the inner product

$$<x[n], y[n]> = \sum_j <\tilde{b}_{jk}, d_{jk}> + <\tilde{a}_{Jk}, c_{Jk}> . \quad (4)$$

Energy equivalence in orthogonal basis space is given as

$$\|x[n]\|^2 = \sum_j \sum_k \|\tilde{b}_{jk}\|^2 + \sum_k \|\tilde{a}_{Jk}\|^2 .$$

For a received vector $y_j[n]$ of length $L$, $<x_j[n+l], y_j[n]>$ is the correlation for shift $l$ of the matching vector $x_j[n]$ in the sampled data domain at the $2^j$ scale. For vectors of finite length, the DWT vectors $\{\tilde{a}_{jk}^{(l)}, \tilde{b}_{jk}^{(l)}\}$ are shift circulant with a period of 2 for one dyadic scale of analysis [23]. Using Mallat's filterbank algorithm [8] to derive DWT coefficients for one dyadic coarser scale, the shift circulant DWT coefficients for even shifts $l = 2p$ are given as [21]

$$\tilde{a}_{(j+1)k}^{e(p)} = \sum_{n=0}^{M-1} \tilde{h}[-n] \tilde{a}_{j(n+2k)}^{(2p)}$$
$$\tilde{b}_{(j+1)k}^{e(p)} = \sum_{n=0}^{M-1} \tilde{g}[-n] \tilde{a}_{j(n+2k)}^{(2p)} .$$

Here $\{\tilde{a}_{(j+1)k}^{e(p)}, \tilde{b}_{(j+1)k}^{e(p)}\}$ represent the DWT vectors $\{\tilde{a}_{(j+1)k}, \tilde{b}_{(j+1)k}\}$ shifted by $p$ samples in left circulant way with even shift $l = 2p$. The lowpass and highpass filters, $\{\tilde{h}[-n], \tilde{g}[-n]\}$ represent the DWT basis vectors to derive $\{\tilde{a}_{(j+1)k}, \tilde{b}_{(j+1)k}\}$ from $\tilde{a}_{jk}$, using Mallat's iterative convolution algorithm. Similarly, for odd shift $l = 2p + 1$,

$$\tilde{a}_{(j+1)k}^{o(p)} = \sum_{n=0}^{M-1} \tilde{h}[-n] \tilde{a}_{j(n+2k)}^{o(2p)}$$
$$\tilde{b}_{(j+1)k}^{o(p)} = \sum_{n=0}^{M-1} \tilde{g}[-n] \tilde{a}_{j(n+2k)}^{o(2p)} ,$$

where $\{\tilde{a}_{jk}^{o}, \tilde{b}_{jk}^{o}\}$ are the DWT vectors of $x_j^o[n] = x_j[n+1]$. A schematic for deriving the left circulant even and odd DWT vectors is shown in Fig. 1.

Circular correlation at $2^j$ scale for all shifts $-L \leq l \leq (L-1)$ in DWT domain is

$$\begin{cases} <x_j[n-L], y_j[n]> = <\tilde{b}_{(j+1)k}^{e(-L/2)}, d_{(j+1)k}> + <\tilde{a}_{(j+1)k}^{e(-L/2)}, c_{(j+1)k}> \\ \vdots \\ <x_j[n], y_j[n]> = <\tilde{b}_{(j+1)k}^{e(0)}, d_{(j+1)k}> + <\tilde{a}_{(j+1)k}^{e(0)}, c_{(j+1)k}> \\ \vdots \\ <x_j[n+L-2], y_j[n]> = <\tilde{b}_{(j+1)k}^{e(L/2-1)}, d_{(j+1)k}> + <\tilde{a}_{(j+1)k}^{e(L/2-1)}, c_{(j+1)k}> \end{cases} \quad (5a)$$

$$\begin{cases} <x_j[n-L+1], y_j[n]> = <\tilde{b}_{(j+1)k}^{o(-L/2)}, d_{(j+1)k}> + <\tilde{a}_{(j+1)k}^{o(-L/2)}, c_{(j+1)k}> \\ \vdots \\ <x_j[n+1], y_j[n]> = <\tilde{b}_{(j+1)k}^{o(0)}, d_{(j+1)k}> + <\tilde{a}_{(j+1)k}^{o(0)}, c_{(j+1)k}> \\ \vdots \\ <x_j[n+L-1], y_j[n]> = <\tilde{b}_{(j+1)k}^{o(L/2-1)}, d_{(j+1)k}> + <\tilde{a}_{(j+1)k}^{o(L/2-1)}, c_{(j+1)k}>. \end{cases} \quad (5b)$$

For even and odd values of $l$, the circular correlation $x_j[n] \circ y_j[n]$ at $2^j$ scale in (5) can be written as

$$x_j[n] \circ y_j[n] = \begin{bmatrix} \tilde{W}_T x_j \\ \tilde{W}_T x_j^o \end{bmatrix} \begin{bmatrix} c_{(j+1)0} \\ \vdots \\ c_{(j+1)(2^{-(j+1)}L)} \\ d_{(j+1)0} \\ \vdots \\ d_{(j+1)(2^{-(j+1)}L)} \end{bmatrix}_{L \times 1} \quad (6)$$

where $\tilde{W}_T x_j$ and $\tilde{W}_T x_j^o$ are the left circulant block matrices, each of size ($L \times L$) produced from (5). $\tilde{W}_T x_j$ and $\tilde{W}_T x_j^o$ together represent SIDWT at the $2^j$ scale as both even and odd indices of lag in $y_j[n]$ are taken care of by the circular shifts of DWT coefficients of $x_j[n+l]$ in the matrices. The correlation $x_j[n] \circ y_j[n]$ in (6) is therefore, shift-invariant.

### III. LIFTING SCHEME FOR DWT FILTERBANKS IN DUAL ANALYSIS BASIS SPACES

*A. Framework for Correlation Using Paraunitary Property of Polyphase Matrices*

The paraunitary property of polyphase matrices for biorthogonal 2 - channel filterbanks is crucial to the lifting factorization of DWT and perfect reconstruction (PR) for analysis and synthesis in dual basis spaces [9]-[11]. Biorthogonal filterbanks need a lowpass FIR filter pair $\{\tilde{h}[-n], h[n]\}$ to generate the highpass filters $\{\tilde{g}[-n], g[n]\}$. PR requires $\{\tilde{h}[-n], h[n]\}$ to be symmetric, i.e. $h[-n] = h[n]$. If the lengths of $\tilde{h}[-n]$ and $h[n]$ are odd, then they must differ by an odd multiple of two [2].

The highpass filters in dual basis spaces are derived from the lowpass filters for linear phase PR [1], [11]:

$$\tilde{g}(z) = z^{-1}h(-z^{-1}), \quad \tilde{h}(z) = -z^{-1}g(-z^{-1}). \quad (7)$$

The polyphase representation expresses the z-domain polynomials of signals and filters in terms of their even and odd phases. For example, the two phases of $x_j[n]$ are given by $x_j(z) = x_{ej}(z^2) + z^{-1}x_{oj}(z^2)$, where $x_{ej}[n] = x_j[2n]$ and $x_{oj}[n] = x_j[2n+1]$ are the even and odd sample sequences of $x_j[n]$ respectively. The polyphase matrices for synthesis and analysis filterbanks are, respectively,

$$P(z) = \begin{bmatrix} h_e(z) & g_e(z) \\ h_o(z) & g_o(z) \end{bmatrix} \quad (8)$$

and

$$\tilde{P}(z^{-1}) = \begin{bmatrix} \tilde{h}_e(z^{-1}) & \tilde{h}_o(z^{-1}) \\ \tilde{g}_e(z^{-1}) & \tilde{g}_o(z^{-1}) \end{bmatrix}. \quad (9)$$

From (7),

$$\tilde{h}_e(z) = g_o(z^{-1}), \; \tilde{h}_o(z) = -g_e(z^{-1}), \; \tilde{g}_e(z) = -h_o(z^{-1}), \; \tilde{g}_o(z) = h_e(z^{-1}). \quad (10)$$

Hence

$$P(z) = \begin{bmatrix} \tilde{g}_o(z^{-1}) & -\tilde{h}_o(z^{-1}) \\ -\tilde{g}_e(z^{-1}) & \tilde{h}_e(z^{-1}) \end{bmatrix}.$$

The determinants of $P(z)$ and $\tilde{P}(z^{-1})$ must be unity or a monomial to satisfy linear phase PR condition [26]. In such a case

$$\tilde{P}(z^{-1})P(z) = I, \quad (11)$$

where $I$ denotes the identity matrix. Equation (11) shows the paraunitary relationship between $P(z)$ and $\tilde{P}(z^{-1})$.

From (8)

$$P^T(z^{-1}) = \begin{bmatrix} h_e(z^{-1}) & h_o(z^{-1}) \\ g_e(z^{-1}) & g_o(z^{-1}) \end{bmatrix} = \begin{bmatrix} \tilde{g}_o(z) & -\tilde{g}_e(z) \\ -\tilde{h}_o(z) & \tilde{h}_e(z) \end{bmatrix} \quad (12)$$

is the Hermitian transpose of $P(z)$.

From (9) and (12) it can be shown that

$$P^T(z^{-1})\tilde{P}^T(z) = I. \quad (13)$$

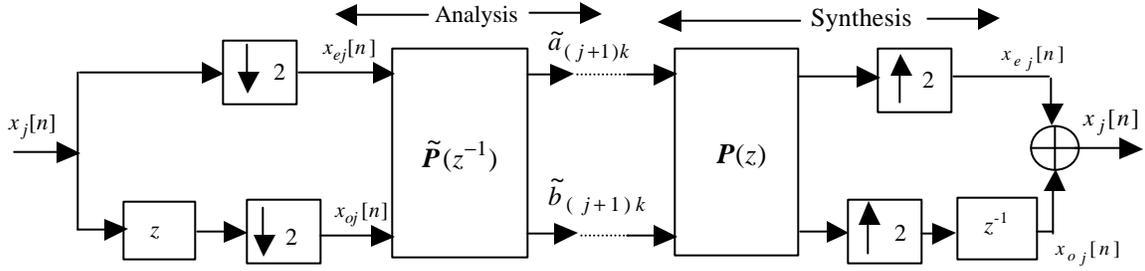

Fig. 2. Polyphase domain representation of DWT in biorthogonal filterbanks.

Comparing (11) and (13), $P^T(z^{-1})$ is the dual analysis matrix derived from $P(z)$. The polyphase representation of 2-channel biorthogonal analysis-synthesis filterbank pair is shown in Fig. 2. This dual analysis formulation in polyphase domain marks a significant departure from the analysis-synthesis approach used in compression, and enables the application of the lifting scheme to SIDWT based correlation.

The outputs of 2-channel analysis filterbank in Fig. 2 represent the DWT coefficients

$$\begin{bmatrix} \tilde{a}_{(j+1)k}(z) \\ \tilde{b}_{(j+1)k}(z) \end{bmatrix} = \tilde{P}(z^{-1}) \begin{bmatrix} x_{ej}(z) \\ x_{oj}(z) \end{bmatrix} \quad (14)$$

Using the paraunitary matrices in dual analysis bases developed in (13) the DWT domain correlation for shift $l$ in (5) can be expressed as

$$\begin{aligned} <x_j[n+l], y_j[n]> &= [\tilde{a}^l_{(j+1)k} \; \tilde{b}^l_{(j+1)k}] \begin{bmatrix} c_{(j+1)k} \\ d_{(j+1)k} \end{bmatrix} \\ &= z^{-1}\left([x^l_{ej}(z^{-1}) \; x^l_{oj}(z^{-1})] \tilde{P}^T(z) P^T(z^{-1}) \begin{bmatrix} y_{ej}(z) \\ y_{oj}(z) \end{bmatrix}\right) \\ &= z^{-1} <x^l_j(z), y_j(z)> \end{aligned} \quad (15)$$

where the operator $z^{-1}$ indicates inverse $z$-transformation of the DWT vectors.

*B. Lifting Scheme for Causal Orthogonal Daubechies Wavelet Filters*

Existing lifting schemes for orthogonal Daubechies wavelet filters are generally noncausal [11], [24]. For radar signal analysis, causal lifting steps is advantageous as it requires less memory and fewer memory calls [24]; moreover, correlation by SIDWT in orthogonal filterbanks is strictly energy-preserving [22]. The filter pair $\{\tilde{h}[-n], \tilde{g}[-n]\}$ is said to be *complimentary* for paraunitary $\tilde{P}(z^{-1})$, and its dual pair $\{h[n], g[n]\}$ at the synthesis end, therefore, is also complimentary. It is shown in [11] that the filterbank operations in such cases are further simplified by factorizing the polyphase matrices in a sequence of alternate upper and lower triangular matrices. By the Euclidean algorithm of iterated divisions, $\tilde{P}(z^{-1})$ can be factorized into Laurent polynomials as quotient terms. One form of this factorization is [9]

$$\tilde{P}(z^{-1}) = \begin{bmatrix} K & 0 \\ 0 & 1/K \end{bmatrix} \left( \prod_{i=0}^{(p-1)} \begin{bmatrix} 1 & 0 \\ Q_{2i+1}(z) & 1 \end{bmatrix} \begin{bmatrix} 1 & Q_{2i}(z) \\ 0 & 1 \end{bmatrix} \right)$$

where $p$ is the even degree of the polynomial $\tilde{h}(z^{-1})$. $Q_{2i+1}(z)$ is the quotient polynomial at an odd iteration number, called the *dual* lifting step, predicting the intermediate odd phase output from the intermediate even phase. At an even-number iteration, $Q_{2i}$ is the quotient polynomial, called the *primal* lifting step that updates the intermediate even phase output from the intermediate odd phase. The scale factor $K$ is the final greatest common divisor (gcd) obtained after iterated divisions.

In terms of the lifting steps factors, the DWT output from analysis filterbank in (14) is

$$\begin{bmatrix} \tilde{a}_{(j+1)k}(z) \\ \tilde{b}_{(j+1)k}(z) \end{bmatrix} = \begin{bmatrix} K & 0 \\ 0 & 1/K \end{bmatrix} \left( \prod_{i=0}^{(p-1)} \begin{bmatrix} 1 & 0 \\ Q_{2i+1}(z) & 1 \end{bmatrix} \begin{bmatrix} 1 & Q_{2i}(z) \\ 0 & 1 \end{bmatrix} \right) \begin{bmatrix} x_{ej}(z) \\ x_{oj}(z) \end{bmatrix}. \quad (16)$$

We now show a causal representation of the lifting scheme for orthogonal wavelet filters for which $\tilde{h}(z^{-1}) = h(z)$ and $\tilde{g}(z^{-1}) = g(z)$. For the Daubechies4 filterbank pair $\tilde{h}(z^{-1}) = h_0 + h_1 z^{-1} + h_2 z^{-2} + h_3 z^{-3}$, $\tilde{g}(z^{-1}) = h_0 z^{-1} - h_1 + h_2 z^1 - h_3 z^2$ are the analysis polynomials [11].

Let $\tilde{g}'(z^{-1}) = z^{-2} \tilde{g}(z^{-1}) = -h_3 + h_2 z^{-1} - h_1 z^{-2} + h_0 z^{-3}$ be the causal representation of $\tilde{g}(z^{-1})$. The new analysis polyphase matrix is

$$\tilde{P}'(z^{-1}) = \begin{bmatrix} h_0 + h_2 z^{-1} & h_1 + h_3 z^{-1} \\ -h_3 - h_1 z^{-1} & h_2 + h_0 z^{-1} \end{bmatrix}$$

Factorization of $\tilde{P}'(z^{-1})$ in lifting steps leads to

$$\tilde{P}(z^{-1}) = \begin{bmatrix} \frac{\sqrt{3}+1}{\sqrt{2}} & 0 \\ 0 & \frac{\sqrt{3}-1}{\sqrt{2}} \end{bmatrix} \left( \begin{bmatrix} 1 & 0 \\ 1 & z^{-1} \end{bmatrix} \begin{bmatrix} 1 & \frac{\sqrt{3}}{4} + \frac{\sqrt{3}-2}{4} z^{-1} \\ 0 & 1 \end{bmatrix} \begin{bmatrix} 1 & 0 \\ -\sqrt{3} & 1 \end{bmatrix} \right). \quad (17)$$

The determinant of $\tilde{P}'(z^{-1})$ is $z^{-1}$. This being a monomial,

the paraunitary property is preserved in $\tilde{P}'(z^{-1})$ by introducing 2-step delay in the highpass analysis filter. The lifting scheme factors for the orthogonal synthesis polyphase matrix in Fig. 2 is derived from (17) as

$$P(z) = \left( \begin{bmatrix} 1 & 0 \\ \sqrt{3} & 1 \end{bmatrix} \begin{bmatrix} 1 & -(\frac{\sqrt{3}}{4}+\frac{\sqrt{3}-2}{4}z^{-1}) \\ 0 & 1 \end{bmatrix} \begin{bmatrix} z^{-1} & 0 \\ -1 & 1 \end{bmatrix} \right) \begin{bmatrix} \frac{\sqrt{3}-1}{\sqrt{2}} & 0 \\ 0 & \frac{\sqrt{3}+1}{\sqrt{2}} \end{bmatrix}.$$

In this case, $P(z)\tilde{P}'(z^{-1}) = z^{-1}I$ is the condition for forward and inverse DWT (IDWT) by causal lifting of Daubechies4 wavelet filters. The IDWT output from synthesis pair of filters in Fig. 2 after interpolation is

$$z^{-2}\begin{bmatrix} x_{ej}(z) \\ x_{oj}(z) \end{bmatrix} = P(z^2)\begin{bmatrix} \tilde{a}_{(j+1)k}(z^2) \\ \tilde{b}_{(j+1)k}(z^2) \end{bmatrix}.$$

Since the dual basis spaces are equal for orthogonal DWT, $\tilde{P}'(z^{-1})$ in (17) can be used for generating DWT coefficients $\{c_{(j+1)k}, d_{(j+1)k}\}$ of $y_j[n]$ also. Therefore, correlation of two vectors by SIDWT at $2^j$ scale as in (6) and (15) can be implemented by causal lifting of orthogonal Daubechies wavelets.

*C. Lifting Scheme for Biorthogonal Wavelet Filters in Dual Analysis Basis Spaces*

The determinant of $\tilde{P}(z^{-1})$ or $P(z)$ remains unity for biorthogonal wavelet filters, permitting new forms of quotient polynomials in (16). Here we exploit this feature by applying the lifting scheme to biorthogonal filters in dual analysis bases as required for SIDWT based correlation. We focus on the (5/3) and (9/7) wavelet filters as these are standard for JPEG2000.

*Case I*: Consider the spline (5/3) biorthogonal filter by Le Gall and Tabatabai [27]. The high pass filters are derived from symmetric $\{\tilde{h}[-n], h[n]\}$, each with two vanishing moments:

$$\tilde{h}(z^{-1}) = -\frac{1}{8}z^2 + \frac{1}{4}z + \frac{3}{4} + \frac{1}{4}z^{-1} - \frac{1}{8}z^{-2},$$

$$\tilde{g}(z^{-1}) = -\frac{1}{4}z^2 + \frac{1}{2}z - \frac{1}{4}.$$

$\tilde{P}(z^{-1})$ is factorized as

$$\tilde{P}(z^{-1}) = \begin{bmatrix} 1 & 0 \\ 0 & \frac{1}{2} \end{bmatrix} \left( \begin{bmatrix} 1 & \frac{1}{4}(1+z^{-1}) \\ 0 & 1 \end{bmatrix} \begin{bmatrix} 1 & 0 \\ -\frac{1}{2}(1+z) & 1 \end{bmatrix} \right).$$

$$h(z) = \frac{1}{4}z + \frac{1}{2} + \frac{1}{4}z^{-1},$$

$$g(z) = -\frac{1}{8}z - \frac{1}{4} + \frac{3}{4}z^{-1} - \frac{1}{4}z^{-2} - \frac{1}{8}z^{-3}.$$

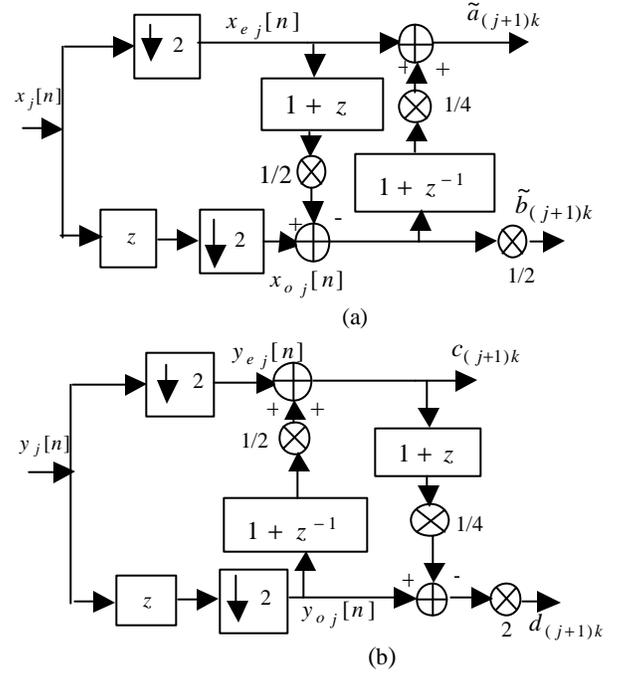

Fig. 3(a), (b). DWT of $\{x_j[n], y_j[n]\}$ by the lifting scheme in dual analysis basis spaces for biorthogonal (5/3) filterbank.

The product of $\tilde{P}(z^{-1})$ with corresponding $P(z)$ is $\tilde{P}(z^{-1})P(z) = 0.5I$. $P(z)$ is therefore scaled by 2 and its lifting steps factors are given as

$$P(z) = \left( \begin{bmatrix} 1 & 0 \\ \frac{1}{2}(1+z) & 1 \end{bmatrix} \begin{bmatrix} 1 & -\frac{1}{4}(1+z^{-1}) \\ 0 & 1 \end{bmatrix} \right) \begin{bmatrix} 1 & 0 \\ 0 & 2 \end{bmatrix}.$$

The dual analysis matrix of $P(z)$ is

$$P^T(z^{-1}) = \begin{bmatrix} 1 & 0 \\ 0 & 2 \end{bmatrix} \left( \begin{bmatrix} 1 & 0 \\ -\frac{1}{4}(1+z) & 1 \end{bmatrix} \begin{bmatrix} 1 & \frac{1}{2}(1+z^{-1}) \\ 0 & 1 \end{bmatrix} \right).$$

Lifting scheme factorization of (5/3) biorthogonal filters in dual analysis bases can, therefore, yield the inner product of vectors in DWT domain as shown in (15). The DWT by the lifting scheme in dual analysis filterbank is shown in Fig. 3(a) and 3(b) for vectors $\{x_j[n], y_j[n]\}$ respectively.

*Case II*: Consider CDF (9/7) biorthogonal filters by Cohen, Daubechies and Feauveau [28]. Both the highpass filters are with four vanishing moments. The original filterbank [28] coefficients are irrational, resulting in irrational lifting parameters that require floating point implementation. Finite precision approximations of multiplier constants are derived recently [7], [10] leading to image compression by the lifting scheme of (9/7) filterbank with insignificant loss in image quality. It has been shown in [10] that finite precision approximations of lifting parameters are possible by introducing a free variable *t* lying between 0.780 and 1.852, with the multiplier constants derived as

$$a = \frac{-2t+1}{4(t-1)}, \quad b = -(t-1)^2, \quad g = \frac{1}{4t(t-1)}, \quad d = t^3 - \frac{7}{4}t^2 + t,$$

and $x = \frac{\sqrt{2}}{t}$.

A rational value $t = (5/4)$ leads to the multiplier constants given in Table I which are corrected versions from [7], [10]. Except the scale factor $x$, the multiplier constants are represented by finite precision numbers. The same lifting parameters are reported in [7], although by a different route of analysis. With these approximations, the finite precision coefficients for (9/7) lowpass filterbanks are

$$\tilde{h}[-n] = \frac{1}{10}\left\{\frac{9}{16}, -\frac{6}{16}, -\frac{24}{16}, \frac{86}{16}, \frac{190}{16}\right\}\frac{1}{\sqrt{2}}, \quad -4 \le n \le 0$$

$$h[n] = \left\{-\frac{3}{64}, -\frac{1}{32}, \frac{19}{64}, \frac{18}{32}\right\}\sqrt{2}, \quad 0 \le n \le 3.$$

The respective normalizing factors $\{1/\sqrt{2}, \sqrt{2}\}$ in $\{\tilde{h}[-n], h[n]\}$ are necessary for approximation to the coefficients given in [28].

TABLE I

RATIONAL LIFTING PARAMETERS FOR (9/7) FILTERBANKS

| a | b | g | d | x |
|---|---|---|---|---|
| $-3/2$ | $-1/16$ | $4/5$ | $15/32$ | $4\sqrt{2}/5$ |

The polyphase matrix $\tilde{P}(z^{-1})$ for (9/7) filterbank is found from $\tilde{P}(z)$ in [11]:

$$\tilde{P}(z^{-1}) = \begin{bmatrix} x & 0 \\ 0 & 1/x \end{bmatrix}\left(\begin{bmatrix} 1 & d(1+z^{-1}) \\ 0 & 1 \end{bmatrix}\begin{bmatrix} 1 & 0 \\ g(1+z) & 1 \end{bmatrix}\begin{bmatrix} 1 & b(1+z^{-1}) \\ 0 & 1 \end{bmatrix}\begin{bmatrix} 1 & 0 \\ a(1+z) & 1 \end{bmatrix}\right).$$

From the derivation of $P^T(z^{-1})$ in (12), the lifting scheme factorization in dual analysis basis space is

$$P^T(z^{-1}) = \begin{bmatrix} 1/x & 0 \\ 0 & x \end{bmatrix}\left(\begin{bmatrix} 1 & 0 \\ -d(1+z) & 1 \end{bmatrix}\begin{bmatrix} 1 & -g(1+z^{-1}) \\ 0 & 1 \end{bmatrix}\begin{bmatrix} 1 & 0 \\ -b(1+z) & 1 \end{bmatrix}\begin{bmatrix} 1 & -a(1+z^{-1}) \\ 0 & 1 \end{bmatrix}\right).$$

DWT of vectors $\{x_j[n], y_j[n]\}$ by the lifting scheme factorization of $\tilde{P}(z^{-1})$ and $P^T(z^{-1})$ are shown in Fig. 4(a), (b).

This case shows that the benefit of robustness against approximation of lifting parameters, demonstrated for compression applications, is also preserved for SIDWT correlation of signal vectors. Since the lifting factorization has been shown here to be applicable for correlation in a wide sense, its benefits are also realizable in the important case of symmetric data and filtering.

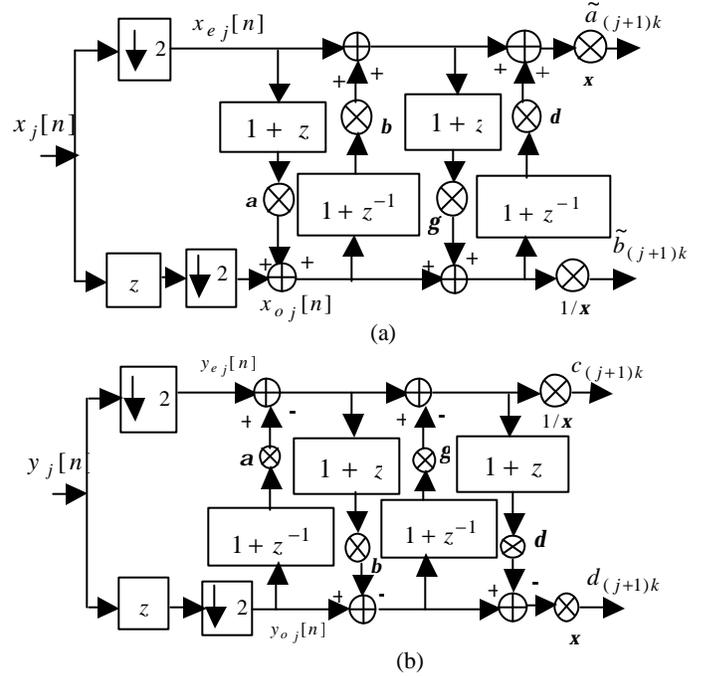

Fig.4(a),(b). DWT of $\{x_j[n], y_j[n]\}$ by lifting scheme factorization of biorthogonal (9/7) filterbank in dual analysis basis spaces.

IV.   MULTIRESOLUTION CORRELATION ALGORITHM BY SIDWT FOR COMPLEX RADAR SIGNALS

The standard model for complex radar transmission is linear frequency modulated (LFM) chirp, the unique feature of such transmission being time-frequency interlocking in the waveform. The autocorrelation function for LFM chirp is a very narrow pulse of $sinc(.)$ shape [30]. Complex LFM chirp function $\bar{x}(t)$ in continuous time is given by,

$$\bar{x}(t) = exp(ipKt^2)rect(t/T_p), \text{ where}$$
$$rect(t) \equiv \begin{cases} 1, & |t| \le 1/2 \\ 0, & otherwise \end{cases} \quad (18)$$

$\bar{x}(t)$ has a bandwidth of $B = KT_p$ sampled over the pulse duration $T_p$ at the rate of $B$ samples/sec to produce a discrete, complex vector $\bar{x}[n]$ of length $BT_p$; the main lobe of the autocorrelation function of $\bar{x}[n]$ is of width $1/B$, the time resolution at original scale and the correlation gain is $\sqrt{BT_p}$. Time resolution of autocorrelation of $\bar{x}_j[n]$ at dyadic coarser scales is increased as $2^j/B$ ($0 \le j \le J-1$), whereas the bandwidth of the coarse resolution signal $\bar{x}_j[n]$ decreases as $B/2^j$. Complex wavelets and complex 2 - channel filterbanks with orthogonality and symmetry properties have been developed recently for complex signal processing [32], [33]. However, analytical DWT for the complex LFM chirp in (18) is not derived here as CWT in [19], [31] or by 2 - channel complex wavelets [33]. Instead, we apply the standard result of linear system theory for deriving analytical wavelet transform $w_T \bar{x}(t)$ by real wavelets only.

$$w_T \bar{x}(t) = \int_{-\infty}^{\infty} \bar{x}(t) \frac{1}{\sqrt{2^j}} \bar{y}^* \left( \frac{t - 2^j k}{2^j} \right) dt$$
$$= <\bar{y}_{j,k}(t), \bar{x}(t)>$$

where $\bar{y}_{j,k}(t)$ is the Hilbert transform of real, continuous wavelet basis function $y_{j,k}(t)$. One standard result of analytical impulse response function is that the analytical system output is twice of the output from real impulse response function for the same analytical input signal [30]. Thus,

$$w_T \bar{x}(t) = 2 <y_{j,k}(t), \bar{x}(t)>$$
$$or,\ w_T \bar{x}[n] = 2 \bar{x}[n] \otimes Re(\bar{y}^*_{j,k}[-n]). \quad (19)$$

Here $\bar{y}^*_{j,k}[-n]$ is the time-reversed, complex conjugate of analytic wavelet basis vector $\bar{y}_{j,k}[n]$. Inner product of two real vectors at $2^j$ scale in (15) is here extended to multiresolution correlation of complex radar signals at dyadic coarser scales of approximation. For three dyadic scales of observation, SIDWT for the real part of $\bar{x}[n]$ is shown in Fig. 5. The SIDWT of Fig. 1 is realized here by lifting with real parameters as shown in Fig. 3(a) or Fig. 4(a).

The periodic shift circulant DWT vector $\{\tilde{a}_{1k}^{e(p)}, \tilde{b}_{1k}^{e(p)}\}$ is derived from $x[n]$, the odd shifted DWT vector $\{\tilde{a}_{1k}^{o(p)}, \tilde{b}_{1k}^{o(p)}\}$ is derived from $x^o[n]$ for correlation at $j = 0$ level. The block DWT matrices $\tilde{W}_T x$ and $\tilde{W}_T x^o$ are correlated with the DWT vector $\{c_{1k}, d_{1k}\}$ derived from $y[n]$ as shown in Fig. 3(b) or Fig. 4(b). There are four block circulant DWT matrices:

$$\{\tilde{W}_T \tilde{a}_{1k}^{e(p)}, \tilde{W}_T \tilde{a}_{1k}^{eo(p)}, \tilde{W}_T \tilde{a}_{1k}^{o(p)}, \tilde{W}_T \tilde{a}_{1k}^{oo(p)}\}$$

each being of size $(L/2 \times L/2)$ at the next coarser level $j = 1$. Therefore, there is a redundancy of $2^{j+1}$ lifting factorization for SIDWT computation at $2^{j+1}$ scale. DWT of $y_1[n]$ of dimension $(L/2 \times 1)$ is derived recursively from $c_{1k}$. At the coarsest level $j = J - 1 = 2$, there are $2^J$ block circlant matrices of only approximation DWT coefficients of $x_{J-1}[n]$, each of dimensions $(L/2^{J-1} \times L/2^{J-1})$. Correlation of the vectors by SIDWT at $0 \le j \le 1$ levels are represented as

$$x_j[n] \circ y_j[n] = \begin{bmatrix} \tilde{W}_T x & \tilde{W}_T \tilde{a}_{1k}^{e(p)} \\ & \tilde{W}_T \tilde{a}_{1k}^{eo(p)} \\ & \tilde{W}_T \tilde{a}_{1k}^{o(p)} \\ \tilde{W}_T x^o & \tilde{W}_T \tilde{a}_{1k}^{oo(p)} \end{bmatrix} \begin{bmatrix} c_{(j+1)k} \\ d_{(j+1)k} \end{bmatrix}. \quad (20)$$

$$j = 0 \quad j = 1$$

Correlation at the coarsest scale is given by,

$$x_j[n] \circ y_j[n] = \begin{bmatrix} \tilde{a}_{2k}^{e(p/2)} \\ \tilde{a}_{2k}^{eo(p/2)} \\ \tilde{a}_{2k}^{eoe(p/2)} \\ \tilde{a}_{2k}^{eoo(p/2)} \\ \tilde{a}_{2k}^{o(p/2)} \\ \tilde{a}_{2k}^{oo(p/2)} \\ \tilde{a}_{2k}^{ooe(p/2)} \\ \tilde{a}_{2k}^{ooo(p/2)} \end{bmatrix} \begin{bmatrix} c_{20} \\ \vdots \\ \vdots \\ c_{2(2^{-2}L)} \end{bmatrix}_{L/4 \times 1}. \quad (20a)$$

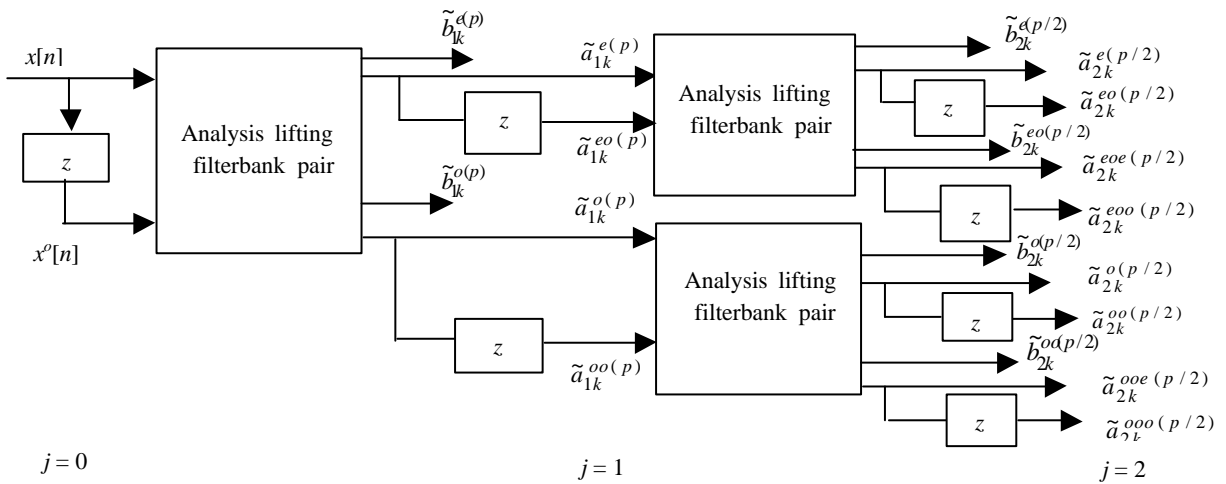

Fig. 5. SIDWT decomposition for real part of $\bar{x}[n]$ for three scales, $0 \le j \le 2$.

Since correlation is shift-invariant for even or odd shifts in $y_j[n]$ at each scale in (20)-(20a), the sample index $l$ for maximum matching between $y_j[n]$ and $x_j[n+l]$ remain independent of scale of observation. It is evident from (19) that SIDWT of $\bar{x}[n]$ is the linear sum of SIDWT for real and imaginary parts of $\bar{x}[n]$, hence the same exercise as in Fig. 5 and in (20)-(20a) are to be performed for imaginary parts of $\bar{x}[n]$ and $\bar{y}[n]$. The shift of peak response of multiresolution correlation of complex signals is scale-invariant, which is essential for MRA of target signature in radar matched filtering. Another interesting feature in (20-20a) is that MRA of complex received signals are done before matched filtering. Therefore, it is possible to derive the coarse resolution MRA first and if required, fine resolution target signature analysis is obtained from coarser resolution analysis in the reverse direction. Inverse lifting steps synthesize the coarser resolution DWT vector at fine scale and follow from IDWT in biorthogonal filterbanks in Fig. 2. The fine scale DWT vector $\{c_{(J-1)k}, d_{(J-1)k}\}$ is derived by lifting steps in Fig. 2 as

$$\begin{bmatrix} c_{(J-1)k}(z) \\ d_{(J-1)k}(z) \end{bmatrix} = \tilde{\boldsymbol{P}}^T(z^2) \begin{bmatrix} c_{Jk}(z^2) \\ d_{Jk}(z^2) \end{bmatrix} \quad (21)$$

## V. RESULTS AND DISCUSSION

### A. Results

Target backscatter is modeled here as the convolution of reflectivity $\boldsymbol{s}_k$ from $P$ number of point targets in the duration of reception, $\bar{y}[n] = \sum_{k=1}^{P} \boldsymbol{s}_k \bar{x}[n-k]$. The analytic received signal is sampled at the rate of $B$ samples/sec to produce $L_w = 2^J$ number of complex samples in $T_w$ sec. A simulation of real part of $\bar{y}[n]$ for target backscatter from five targets is shown in Fig. 6(a).

Matched filter output at three dyadic coarser scales for target backscatter in Fig. 6(a) is shown in Fig. 6(b)-(d). Here DWT is performed by lifting of CDF (9/7) filterbank. Although the derivation in (20-20a) is independent of the wavelet filter used, it is difficult to keep track of the coefficients produced using orthogonal wavelets as it results expansion of the DWT coefficients at each scale [22]. Secondly, orthogonal wavelet filters do not maintain linearity in phase that is required for matched filtering of analytical signals at dyadic scales.

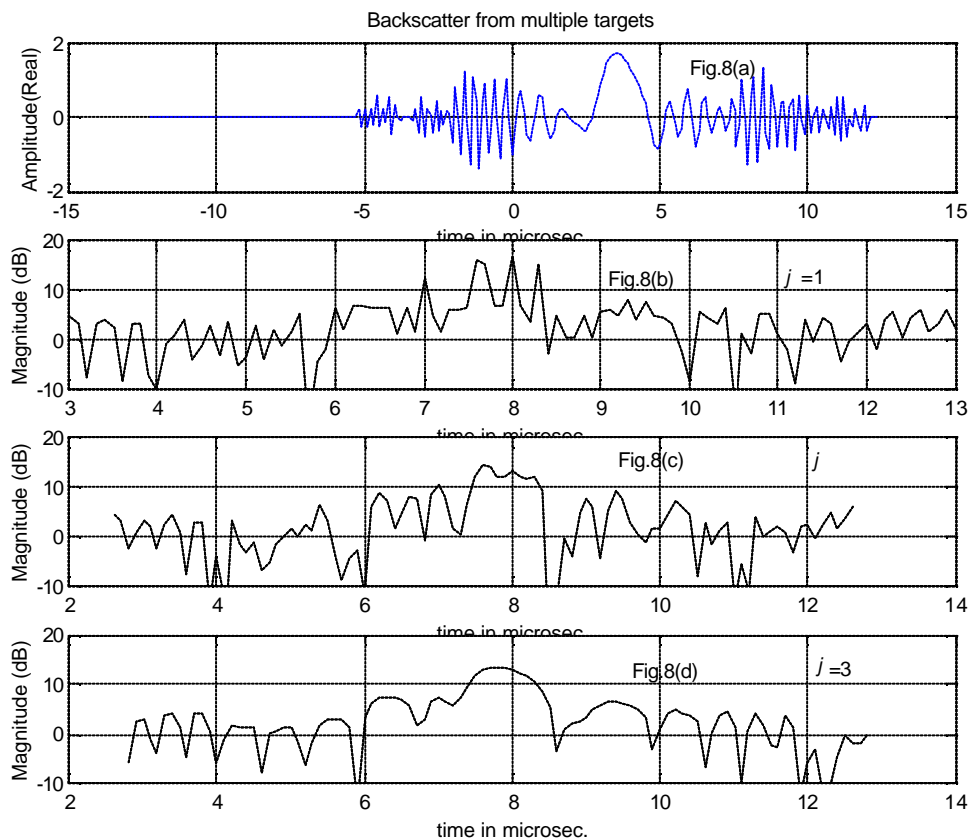

Fig. 6(a). Real part of backscatter $\bar{y}[n]$ from five adjacent point targets, $BT_p = 160$; $L_w = 244$.

Fig. 6(b)-(d). Magnitude(dB) of multiresolution correlation of $\bar{y}_j[n]$ with $\bar{x}_j[n]$ for three dyadic scales, $0 \le j \le 2$ using lifting steps of CDF(9/7) filterbank.

In Fig. 6(b)-(d), magnitude of multiresolution complex correlation $\bar{x}_j[n] \circ \bar{y}_j[n]$ in dB is shown at three dyadic scales. In Fig. 6(b), resolution is not fine enough to separate two very nearby targets, but adjacent three other targets are identified above 10dB magnitude. At the next coarser scale in Fig. 6(c), target reflectivity of four nearby targets merge into two distinct peaks around 8 microsec, but the fifth target at 7 microsec delay is also identifiable with magnitude touching 10dB. Finally, at the coarsest scale in Fig. 6(d), target reflectivity of all the five targets are merged in a single extended peak above 10dB around a delay of 8 microsec. Analysis of such multiresolution target reflectivity is proven to be beneficial [13], [14] for improved declaration of multiple targets in presence of ground clutter. It is to be mentioned that all of the multiresolution target reflectivity in Fig. 6(b)-(d) can be derived simultaneously in a parallel implementation of (20-20a).

*B. Discussion*

Apart from parallelism in the algorithm for multiresolution target detection, another major advantage from the PR property is that the coarse scale correlation can be implemented at a slower throughput independent of and without implementing finer scale correlation. The complex data throughput at the $2^J$ scale is $B/2^J$ and (21) leads to coarse to fine resolution formation hierarchically [34].

Another aspect of parallelism lies in the computation of DWT of $\bar{y}_j[n]$ by lifting. DWT of real part of $\bar{y}_j[n]$ for two dyadic scales is shown in Fig. 7 using lifting of LeGall (5/3) biorthogonal filterbank in Fig. 3(b). It is seen from Fig. 7 that there is one sample delay in generating $c_{1k}$ from $y[n]$ and one sample advance to generate $d_{1k}$. After this, DWT computation at each scale can be done in parallel fashion. Secondly, DWT by lifting steps is in place computation; the intermediate output from the first filter serves as the input of the second filter at each scale. The final output DWT samples can be stored at the memory locations for intermediate values. Thus, DWT by the lifting steps is analogous to parallel, in place FFT computation in butterflies. The difference is that FFT butterfly is a complex operation whereas lifting operates on real data; for DWT of complex data, a pair of filterbank by lifting steps is necessary as explained in Section IV.

We now briefly dwell upon the computational aspect of the algorithm. The number of 2 - channel DWT coefficients for $2^J$ scale decomposition is bound; $(1/2 + 1/4 + \cdots + 1/2^J)L_w + L_w/2^J \approx L_w$. The recursive tree structure of critically sampled DWT has the advantage that the number of computations does not grow with increasing levels of decomposition $J$; in fact it is bound to $2C_0$ [35], where $C_0$ is the number of multiplications and addition ($MUL/ADD$) operations/sample in a single channel of the FIR filterbank in Fig. 1. Estimation of computational load, therefore, depends on different realizations of filterbanks.

From the symmetry property of the biorthogonal filters, $2C_0 = \{(M+1)MUL + (M+3)ADD\}$ for implementation of DWT, $M = \mathbf{max}\{|h[n]|, |g[n]|\} + 1$ being the order of the filter. The primal and dual Laurent polynomials in the paraunitary matrices, respectively $(1+ z^{-1})$ and $(1+ z)$ are symmetric. In that case, the number of multiplications and additions in the lifting factors are minimum [29]. A detailed study of ($MUL/ADD$) operations in the integer format of the lifting scheme is given in [25]; $2C_0 = \{4MUL + 12ADD\}$ for (9/7) filterbank and $2C_0 = \{5ADD\}$ for (5/3) filterbank. It is to be noted that 2 - channel output is *simultaneously* available from the same lifting step.

At each coarser scale of SIDWT decomposition the number of DWT coefficients at each output channel is $2^{J/j}$. There are $2^{j+1}$ channels ($2^j$ pair of filterbanks), the total computational load is $2^{J+1}C_0$ for real DWT coefficients only. The same number of computations is to be done for the imaginary part of the input also. Hence the computational load for SIDWT of complex input vector at each dyadic coarser scale of Fig. 5 is $4L_wC_0$. Apart from the vector multiplications in the block circulant matrices in (20-20a), total computation cost for analytic DWT $\approx 4L_w \log_2(L_w)C_0$ for large $L_w$. The computational load of analytic DWT for realization in symmetric filterbanks and by lifting steps is summarized in Table II.

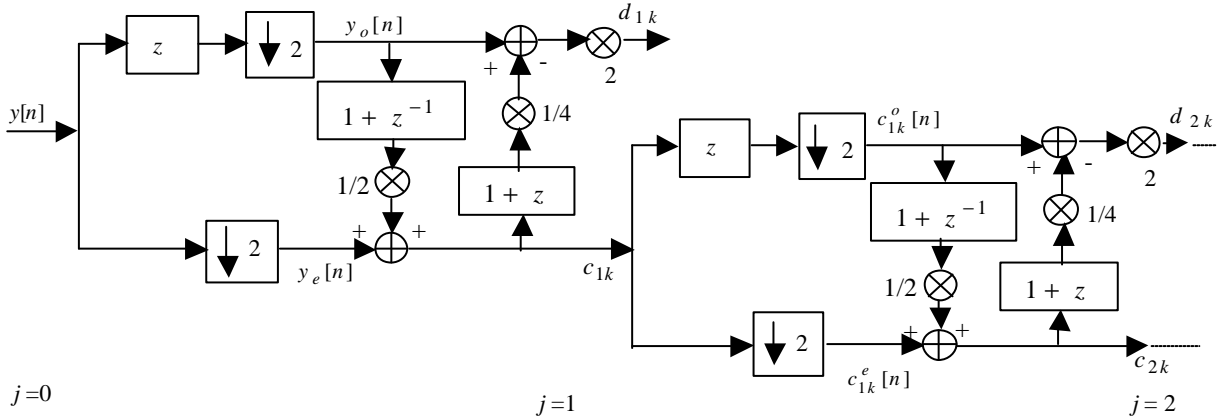

Fig.7. DWT of $y_j[n]$ by lifting steps as shown in Fig. 3(b) for two scales, $1 \le j \le 2$.

TABLE II

COMPUTATIONAL LOAD FOR SIDWT OF ANALYTIC SIGNAL

| Biorthogonal filterbank | Computation Load $4 L_w \log_2(L_w) C_0$ | |
|---|---|---|
| | Symmetric | Lifting |
| CDF(9/7) | $44 L_w \log_2(L_w)$ | $32 L_w \log_2(L_w)$ |
| LeGall(5/3) | $36 L_w \log_2(L_w)$ | $10 L_w \log_2(L_w)$ |

## VI. CONCLUSIONS

This paper has been devoted to developing an analytical framework for the application of the lifting scheme of DWT to signal correlation. Correlation of complex signals is encountered routinely in radar signal processing. The use of DWT enables simultaneous multiresolution correlation which has multiple benefits in the radar context. The lifting scheme is a relatively recent development that is primarily utilized for image compression and is now part of the JPEG 2000 standards. A large body of theory and hardware has been developed for this application. However, these cannot readily be applied to correlation which is performed only in the analysis domain, unlike image compression which has both analysis and synthesis components. An exhaustive analytical formulation enabling adaptation of the lifting scheme of DWT to multiresolution correlation is the primary contribution of this paper.

It is shown analytically that both orthogonal and biorthogonal filterbanks can be factorized in dual analysis basis spaces in the DWT domain. A consistent formalism has been used throughout the formulation to facilitate the proof of the paraunitary property of the dual analysis polyphase matrices. Shift and scale variance, major difficulties in achieving correlation by critically sampled DWT, are overcome by the use of shift circulant property of the DWT coefficients obtained through the lifting formulation. It is demonstrated by simulation results and estimation of computation load that multiresolution correlation of complex radar signals by such lifting steps leads to a practical, fast and parallel algorithm at dyadic coarse scales of observation.

## ACKNOWLEDGMENT

This work was supported by DEAL, Dehradun, India under a CARS project.

## REFERENCES

[1] I. Daubchies, *Ten Lectures on Wavelets*. Philadelphia, PA: SIAM, 1992, ch.5, 8.
[2] M.Vetterli and C.Herley, "Wavelets and filterbanks: theory and design," *IEEE Trans. Signal Processing*, vol.40, no.9, pp.2207-2232, Sept. 1992.
[3] D.S.Taubman and M.W.Marcellin, *JPEG2000, Image Compression Fundamentals, Standards and Practice*, Kluwer Academic Publishers, 2002, ch.6.
[4] K.Andra, C.Chakrabarti and T.Acharya, "A VLSI architecture for lifting-based forward and inverse wavelet transform," *IEEE Trans. Signal Processing*, vol.50, no.4, pp.966-977, April 2002.
[5] W.Jiang and A.Ortega, "Lifting factorization-based discrete wavelet transform architecture design," *IEEE Trans. Circuits and Systems for Video Technology*, vol. 11, no. 5, pp. 651-657, May 2001.
[6] S. Barua, J.E. Carletta, K.A. Kotteri, A.E. Bel, "An efficient architecture for lifting-based two-dimensional discrete wavelet transforms," *INTEGRATION, the VLSI journal*, vol.38, pp.341–352, 2005.
[7] K.A.Kotteri, S.Barua, A.E.Bell and J.E.Carletta, "A comparison of hardware implementations of the biorthogonal 9/7 DWT; convolution versus lifting," *IEEE Trans.Circuits and Systems-II*, vol.52, no.3, pp.256-260, May 2005.
[8] S.G. Mallat, "A theory for multiresolution signal decomposition: the wavelet representation," *IEEE Trans Pattern Anal. Machine Intel.*, vol.11,no.7, pp.674-693,1989.
[9] D.Stefanoiu and I.Tabus, "Euclidean lifting schemes for I2I wavelet transform implementation," *Studies in information and control*, vol.11, no.3, pp.255-270, Sept.2002.
[10] L.Cheng ,D.L.Liang and Z.H.Zhang, " Popular biorthogonal wavelet filters via a lifting scheme and its application in image compression," *IEE Proc. Vision, Image ,Signal Process.*,vol.150,no.4, pp.227-232, Aug.2003.
[11] I.Daubechies and W.Sweldens, " Factoring wavelet transforms into lifting steps," *J.Fourier Analysis and Applications*, vol.4, no.3, pp.245-267, 1998.
[12] W.Sweldens, " The lifting scheme: a construction of second generation wavelets," *SIAM J. Math. Anal.*,vol.29, no.2, pp.511-546,1997.
[13] N.S. Subotic, B.J. Thelen, J.D. Gorman, and M F. Reiley, "Multiresolution detection of coherent radar targets," *IEEE Trans. Image Processing*, vol.6, no.1, pp.21-35, 1997.
[14] W.W.Irving, A.S.Willsky, L.M.Novak, " A multiresolution approach to discriminating targets from clutter in SAR imagery," *Proc. of SPIE* , vol. 2487,pp.272-299.
[15] A. Mojsilovic´, M. V. Popovic´ and D. M. Rackov, " On the selection of an optimal wavelet basis for texture characterization," *IEEE Trans. Image Processing*, vol.9, no.12, pp.2043-2050,Dec.2000.
[16] J.R. Sveinsson and J.A. Benediktsson, "Almost translation invariant wavelet transformations for speckle reduction

of SAR images," *IEEE Trans. Geo Sci. and Remote Sensing*, vol.41, no.10, pp.2404-2408, Oct. 2003.

[17] F. Argenti and L.Alparone, " Speckle removal from SAR images in the undecimated wavelet domain," *IEEE Trans. Geo Sci. and Remote Sensing*, vol.40, no.11, pp.2363-2373, Nov.2002.

[18] H.S. Stone, " Progressive wavelet correlation using Fourier methods," *IEEE Trans. Signal Processing*, vol. 47, no.1, pp.97-107, 1999.

[19] N. Kingsbury, "Shift invariant properties of the dual-tree complex wavelet transform," Proc. IEEE Int. Conf. Acoustics, Speech, and Signal Processing, vol.3, pp.1221 – 1224, March 1999.

[20] J.O.Chapa and R.M.Rao, "Algorithms for designing wavelets to match a specified signal," *IEEE Trans. Signal Processing*, vol.48, no.12, pp.3395-3406,Dec. 2000.

[21] C.Bhattacharya and A.Kar, "Multiple scale correlation of signals by shift-invariant discrete wavelet transform," *IEE Proc. Vision, Image, Signal Process.*, vol.152, no.6,pp.837-845,Dec.2005.

[22] B.E.Usevitch, " A tutorial on modern lossy wavelet image compression: foundations of JPEG 2000," *IEEE Signal Processing Magazine*, pp.22-35, September 2001.

[23] G. Beylkin, "On the representation of operators in bases of compactly supported wavelets," *SIAM J. Numer. Anal.*, vol. 6, no.6, pp.1716-1740, 1992 .

[24] C. G. Chrysafis and A.Ortega, " Minimum memory implementations of the lifting scheme," *Proc. SPIE*, vol. 4119, pp.313-324, 2000.

[25] M.D. Admas and F. Kossentini, "Reversible integer-to-integer wavelet transforms for image compression: Performance evaluation and analysis," *IEEE Trans. Image Processing*, vol. 9,no.10, pp.1010-1024, June 2000.

[26] G. Strang and T. Nguyen, *Wavelet and Filter Banks* (pp.141), Wellesley-Cammbridge Press.

[27] D. Le Gall and A. Tabatabai, " Sub-band coding of digital images using symmetric short kernel filters and arithmetic coding techniques," in *Proc. IEEE Int. Conf. Acoustics, Speech, Signal Processing*, vol.2, pp.761-764, Apr.1988.

[28] A. Cohen, I. Daubechies and J. Feauveau, "Bi-orthogonal bases of compactly supported wavelets," *Comm. Pure, Appl. Math.*, vol.45, pp.485-560,1992.

[29] Yan-Kui Sun, "Symmetric lifting factorization and matrix representation of biorthogonal wavelet transforms," *International J. Wavelets, Multiresolution and Info. Processing*, vol. 1, no. 4, pp.465-479, 2003.

[30] P.Z. Peebles, Jr., *Radar principles,* John Wiley & Sons, 1998.

[31] J. Magarey and N. Kingsbury "Motion estimation using a complex-valued wavelet transform," *IEEE Trans. Signal Processing*, vol. 46**,** no. 4, pp.1069-1084, 1998.

[32] P.L.Shui, Z.Bao, Y.Y.Tang "Three band biorthogonal interpolating complex wavelets with stopband suppression via lifting scheme," *IEEE Trans. Signal Processing*, vol. 51, no.3, pp.1293-1305, May 2003.

[33] X.Q.Gao, T.Q. Nguyen and G. Strang, "A study of two-channel complex-valued filterbanks and wavelets with orthogonality and symmetry properties," *IEEE Trans. Signal Processing*, vol.50, no.4, pp.824-833, April 2002.

[34] C. Bhattacharya, J.Roy and A. Kar, "Speckle reduction and hierarchical resolution formation in SAR signal domain," 5th *Int'l Conf. Adv. Pattern Recog.*, ISI, Kolkata, India, 13-16th Dec., 2003.

[35] O. Rioul and P. Duhamel, "Fast algorithms for discrete and continuous wavelet transforms," *IEEE Trans. Information Theory*, vol.38, no.2, pp. 569-586,March 1992.